\documentclass[letter,twocolumn]{jpsj2}

\title{Evaluation of Exchange Interactions of the Spin Doughnut $\mathrm{Mo_{75}V_{20}}$}
\author{Shunsuke {\sc Takemura}\thanks{Present address: Earthquake Research Institute, University of Tokyo, Tokyo.}
and Yoshiyuki {\sc Fukumoto}\thanks{E-mail address: yfuku@ph.noda.tus.ac.jp}}
\inst{Department of Physics, Faculty of Science and Technology, Tokyo University of Science, Noda, Chiba 278-8510}
\recdate{\hspace{3cm}}

\abst{
Magnetic properties of the spin doughnut $\mathrm{Mo_{75}V_{20}}$, which is a frustrated magnetic cluster with twenty spins, are studied by using the finite-temperature Lanczos method.
Our model Hamiltonian consists of one ten-spin ring and two five-spin rings with nearest-neighbor antiferromagnetic couplings $J$ and $J^{\prime\prime}$.
The five-spin rings are coupled to the ten-spin ring by an antiferromagnetic exchange interaction $J'$ so as to form isosceles triangles of one $J$-bond and two $J^{\prime}$-bonds.
Fitting the theoretical magnetic susceptibility with the experimental one measured by M\"{u}ller {\it et al.},
the exchange parameters are estimated as $J=388$ K, $J^{\prime}=163$ K and $J^{\prime\prime}=81$ K.
The most interesting feature of this system is that the two five-spin rings behave like two free-spins at low temperatures in despite of $J^{\prime}>J^{\prime\prime}$.
We also calculate temperature dependence of the specific heat, and find three peaks at $T\sim J$, $J^{\prime\prime}$ and $J^{\prime}/100$.
}

\kword{quantum spin icosidodecahedron, sawtooth lattice, $\mathrm{Mo_{75}V_{20}}$, 
Heisenberg antiferromagnet, spin susceptibility, specific heat, finite-temperature Lanczos method}

\begin{document}
\maketitle


A molecular quantum-spin icosidodecahedron was found to be realized in a kind of giant molybdenum-oxide-based spherical clusters,
and has provided a new stage of the investigation of frustrated spin systems.
The term {\it icosidodecahedron} means a polygon formed by twenty triangles and twelve pentagons. 
Typical examples of molecular quantum-spin icosidodecahedrons are $\mathrm{Mo_{72}Fe_{30}}$\cite{Muller0} and $\mathrm{Mo_{72}V_{30}}$,\cite{Muller1,Muller2} 
which are often called {\it spin balls}.

The $S=5/2$ spin ball $\mathrm{Mo_{72}Fe_{30}}$ has been studied extensively by many authors.~\cite{Schnack,Garlea,Exler,Hasegawa,Cepas,Schroder}
The main concern of this material is to study the crossover between quantum and classical regions.
In addition, the antiferromagnetic interaction, $J\sim 1.6$ K, is so small that one can saturate the magnetization experimentally by a magnetic field $\sim 20$ T.
As for the $S=1/2$ spin ball $\mathrm{Mo_{72}V_{30}}$, M\"{u}ller {\it et al.} measured the spin susceptibility,
which asymptotically approaches zero as $T\to 0$, featuring a maximum at 11.0 K.\cite{Botor}
They also analyzed this experimental susceptibility on the basis of the antiferromagnetic Heisenberg model with nearest-neighbor coupling $J$.\cite{Muller3}
By using quantum Monte Carlo method to study the high-temperature behavior for $T>120$ K, 
they obtained $J=245$K and the Land\'{e} factor of $g=1.95$.
However, due to the negative sign problem of the method, 
it is not clear whether the lower-temperature susceptibility can be reproduced within the nearest-neighbor Heisenberg model or not.

Replacing ten $\mathrm{V^{4+}}$ in top and bottom of the spin ball $\mathrm{Mo_{72}V_{30}}$ with neutrality ions,
we obtain the $S=1/2$ spin doughnut $\mathrm{Mo_{75}V_{20}}$,\cite{Muller1,Muller2} which is the main subject in this paper.
The susceptibility $\chi$ of the spin doughnut was measured by M\"{u}ller {\it et al.} from $T=300$ K to 2 K.
A quite interesting behavior was found in the experiment: $\chi(T)$ increases as temperature decreases, and shows no maximum down to 2 K.
Especially, they pointed out that the susceptibility follows the Curie law, $\chi =\mathrm{const}/T$, below 20 K.
The magnitude of the Curie constant is as large as that of two free-spins par spin doughnut.
In this letter, we assume that the low-temperature Curie law is an inherent property of $\mathrm{Mo_{75}V_{20}}$,
and study how the free-spin like behavior emerges.

\begin{figure}[b]
\begin{center}
\includegraphics[width=.8\linewidth]{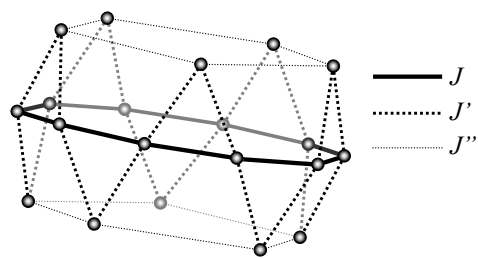}
\caption{Schematic representation of the Heisenberg model defined in eq.~(\ref{hsp:1})
with $L=20$ for the spin doughnut $\mathrm{Mo_{75}V_{20}}$.}
\label{fig:sp4}
\end{center}
\end{figure}

Magnetism of $\mathrm{Mo_{75}V_{20}}$ is expected to be described by the following model Hamiltonian:
\begin{eqnarray}
	\mathcal{H}=&&\hspace{-8mm}J\sum_{i=1}^{L/2}\mib{S}_{2i-1}\cdot \mib{S}_{2i+1}+J^{\prime}\sum_{i=1}^L \mib{S}_{i}\cdot \mib{S}_{i+1}
	\nonumber \\
	&&\hspace{-8mm}+J^{\prime\prime}\sum_{i=1}^{L/2}\mib{S}_{2i}\cdot\mib{S}_{2i+4}
\label{hsp:1}
\end{eqnarray}
with $L=20$ (see Fig.~\ref{fig:sp4}). As seen in Fig.~\ref{fig:sp4}, $J$- and $J^{\prime\prime}$-bonds, respectively,
construct a ten-spin ring and two five-spin rings, 
and a spin on the five-spin ring is connected to two spins on the ten-spin ring by $J^{\prime}$-bonds.

This model was proposed for the first time by M\"{u}ller {\it et el.} in their theoretical analysis of the experimental susceptibility.~\cite{Muller2}
Omitting $J^{\prime\prime}$-bonds first, they calculated temperature dependence of the susceptibility for $L=6$-16 systems, 
which are smaller than $L=20$ due to the limitation of their computer memory,
and found that the best description of the experimental data was obtained for $L=12$.
It was reported that the low temperature Curie law due to two uncorrelated spins was reproduced for a narrow range around $J^{\prime}=0.55 J$.
They finally introduced $J^{\prime\prime}$-bonds to improve the agreement between the theory and experiments, and resulted in $J=288$K,
$J^{\prime}=0.55 J$, $J^{\prime\prime}=0.20 J$ and $g=1.95$.

In this letter, we calculate the eigenvalue distribution function of eq.~(\ref{hsp:1}) with $L=20$
to study the susceptibility and specific heat of the spin doughnut $\mathrm{Mo_{75}V_{20}}$.
We are going to show that the present  model shows the low-temperature Curie law for a wide parameter region, 
which is not restricted to the region around $J^{\prime}=0.55 J$.
The origin of the low-temperature Curie law is that the five-spin rings are in the doublet states at temperatures below $J^{\prime\prime}$.
As described later, our theoretical analysis of the experimental susceptibility of $\mathrm{Mo_{75}V_{20}}$ leads to $J^{\prime}=0.42 J$ and $J^{\prime\prime}=0.21 J$.
Note that $J^{\prime}>J^{\prime\prime}$ is naturally expected from possible exchange paths in $\mathrm{Mo_{75}V_{20}}$.
It should be stressed that, in spite of $J^{\prime}>J^{\prime\prime}$, there appears the doublet behavior of the five-spin rings in thermodynamics.
Because of the zigzag form of $J^{\prime}$-bonds, the five-spin rings are possible to be almost independent from the ten-spin ring in $\mathrm{Mo_{75}V_{20}}$.


We here describe our calculation method. 
The spin doughnut has twenty $S=1/2$ operators, so it is difficult to calculate all eigenvalues by the Householder method.
An alternating way is to calculate the eigenvalue distribution function (EvDF)
\begin{equation}
	\rho(\omega)=\mbox{Tr}\;\delta(\omega-\mathcal{H}) 
\label{me:1}
\end{equation}
by using the finite-temperature Lanczos method.~\cite{Otsuka}
The EvDF can be rewritten as
\begin{eqnarray}
	\rho(\omega)&&\hspace{-7mm} =\sum_{M}\rho_{M}(\omega)\nonumber \\
	&&\hspace{-7mm} = -\frac{1}{\pi}\sum_{M}\sum_{n=1}^{N_{M}} \mbox{Im} \;_{M}\langle n | \mathcal{G}(\omega+i 0)| n \rangle_{M},
\label{me:2}
\end{eqnarray}
where $\{| n \rangle_{M}\}$ denotes the Ising basis with the total $S^z$ of $M$, 
$N_M$ is the dimension of the $M$-subspace,
and $\mathcal{G}(z)=(z-\mathcal{H})^{-1}$ is Green's function.
In the finite-temperature Lanczos method, the sum over $n$ is estimated by sampling random vectors $\{\phi\}$ in the $M$-subspace,
and the matrix elements $_{M}\langle \phi | \mathcal{G}(z)| \phi \rangle_{M}$ are calculated by using the continued fraction form
with Lanczos coefficients of the Hamiltonian matrix.
Once the EvDF is obtained, expectation values of $\mathcal{H}^l$ and $(S_{\rm{tot}}^z)^l$ are calculated via
\begin{equation}
	\langle \mathcal{H}^l \rangle=\frac{1}{Z}\int_{-\infty}^{\infty}d\omega e^{-\beta\omega}\omega^l \rho(\omega),
\end{equation}
\begin{equation}
	 \langle (S_{\rm{tot}}^z)^l \rangle=\frac{1}{Z}\int_{-\infty}^{\infty}d\omega e^{-\beta\omega} \sum_M M^l \rho_M(\omega),
\end{equation}
where $Z=\int_{-\infty}^{\infty}d\omega e^{-\beta\omega}\rho(\omega)$,
and thus we obtain the susceptibility $\chi=L^{-1}[\langle (S_{\rm{tot}}^z)^2 \rangle-\langle S_{\rm{tot}}^z \rangle^2]/T$
and the specific heat $C=L^{-1}[\langle \mathcal{H}^2 \rangle-\langle \mathcal{H} \rangle^2]/T^2$ per spin.
It has been known that this procedure gives fairly good results of thermodynamic quantities.\cite{Otsuka,Takemura}


Before proceeding to our calculated results, let us make a rough estimation of $a\equiv J^{\prime}/J$ and $b\equiv J^{\prime\prime}/J$ for $\mathrm{Mo_{75}V_{20}}$.
We expect $b<a$, because the exchange paths of $J^{\prime\prime}$ are longer than those of $J^{\prime}$.
So we first consider the ground state phase diagram of the sawtooth lattice Heisenberg model with $J$ and $J^{\prime}$, omitting $J^{\prime\prime}$-bonds.
We call spins on tips of the sawtooth ``tip spins" and the others ``bottom spins".
At $a=1$, the ground state is the exact dimer ground state, where  tip and bottom spin pairs are in the singlet dimer state.~\cite{Sen}
By using the level-spectroscopy method,~\cite{Nomura} Blundell and N\"{u}\~{n}ez-Regueiro studied the quantum phase transition point between 
the dimer phase and the spin fluid phase in the region of $a>1$, and showed that the transition point exists at $a=2.052$.~\cite{Blundell}
They also  studied the region of $a<1$, and found that the dimer phase ends at $a=0.653$.~\cite{Blundell}
In the ground state for $a>0.653$, the dimer or spin-fluid state, a tip spin correlates with a bottom spin.
Then, we expect small $b$ term gives just a perturbation to the system, and the low-temperature Curie law is not obtained.
Therefore, in order to get the low-temperature Curie law, we concentrate on the region of $a<\sim0.6$, 
where $b$ terms may play an essential role, not a weak perturbation.
If we suppose the low-temperature Curie law is obtained, the onset temperature should be dominated by the strength of $J^{\prime\prime}$-bonds.
As suggested by the study of the spin ball, we expect the magnitude of $J$ in the spin doughnut is several hundred K.
The experiment on the spin doughnut has indicated that the onset temperature of the low-temperature Curie law is about 20 K,
which leads to $b\sim 0.1$.

\begin{figure}[b]
\begin{center}
\includegraphics[width=.75\linewidth]{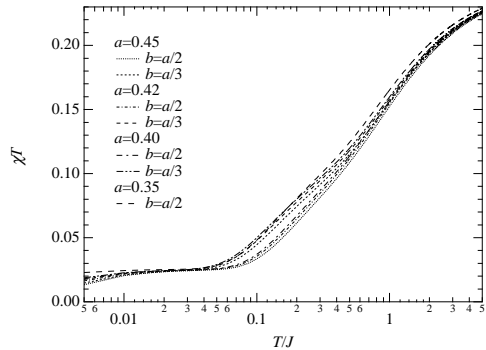}
\caption{Overall behavior of $\chi T$ par spin 
for several values of $a=J^{\prime}/J$ and 
$b=J^{\prime\prime}/J$. A plateau in $\chi T$ at low 
temperatures is seen in all data.}
\label{mag:sp5}
\end{center}
\end{figure}

The crucial point of our scenario is whether the effect of $J^{\prime\prime}$-bonds indeed dominates over that of stronger $J^{\prime}$-bonds.
In order to check this point, we show our calculated results of $\chi T$ per spin for several values of $a$ and $b$ in Fig.~\ref{mag:sp5}.
We find in this figure that the susceptibility shows the Curie law of two-uncorrelated spins, $\chi T=0.025$, 
at low temperatures for all parameter sets used in this calculation.

We turn to analysis of the experimental susceptibility to estimate the exchange parameters for $\mathrm{Mo_{75}V_{20}}$.
We intend to make two types of fittings, which we call ``case 1" and ``case 2", respectively.
In case 1, we use the experimental data of the fresh sample in Fig.~16 of Ref.~2.
The sample is assumed to contain no magnetic impurities, and the Land\'{e} factor is determined by the condition that
the low temperature tail of the experimental data agrees with the Curie law of two uncorrelated-spins.
In case 2, we take into account the existence of magnetic impurities with $S=1/2$ and the Land\'{e} factor is fixed to $g=2$.
When $g=2$ is assumed, the low temperature tails of the fresh sample data and the aged sample data in Fig.~16 of Ref.~2
are somewhat larger than that of the Curie law of two uncorrelated-spins. 
We use those differences to estimate concentrations of the magnetic impurities.

\begin{figure}[b]
\begin{center}
\includegraphics[width=.99\linewidth]{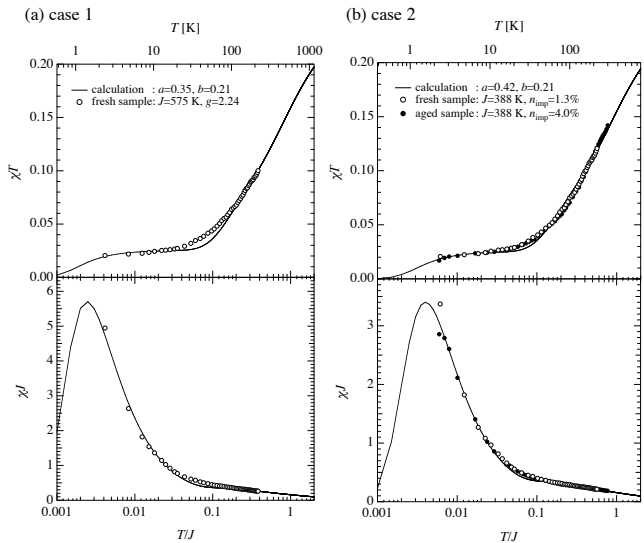}
\caption{Comparison between the calculated and experimental susceptibilities per spin.
In (a) case 1, we assume there are no magnetic impurities. In (b) case 2, 
we fix the Land\'{e} factor as $g=2.00$, and use the impurity concentration, $n_{\rm{imp}}$, 
as a fitting parameter.}
\label{mag:sp6}
\end{center}
\end{figure}

Following the fitting procedure of case 1, we obtain $g=2.24$ and find the best fit is achieved at $J=575$ K, $a=0.35$ ($J^{\prime}=201$ K) and $b=0.21$ ($J^{\prime\prime}=101$ K).
For the spin ball (SB) $\mathrm{Mo_{72}V_{30}}$, the exchange constant $J_{\mathrm{SB}}$ and the Land\'{e} factor $g_{\mathrm{SB}}$ have been estimated as 
$J_{\mathrm{SB}}=245$ K and $g_{\mathrm{SB}}=1.95$, respectively.\cite{Botor}
The discrepancy between those and the present results of $J=575$ K and $g=2.24$ for the spin doughnut may be too large.
In Fig.~\ref{mag:sp6}(a), the calculated results of $\chi T$ and $\chi$ are compared with the experimental result.
We find in this figure that disagreement is pronounced at intermediate temperatures, $0.04<T/J<0.15$. 

When we use the procedure of case 2, the impurity concentrations of the fresh and aged samples are, respectively, estimated as 1.3\% and 4.0\%.
We also found that both samples give almost the same curve, after subtracting the impurity contribution from the raw data of the susceptibility.
Using this susceptibility data, we obtain $J=388$ K, $a=0.42$ ($J^{\prime}=163$ K) and $b=0.21$ ($J^{\prime\prime}=81$ K) via the fitting procedure.
The value of $J=388$ K is closer to $J_{\mathrm{SB}}=245$ K than case 1.
The theoretical and experimental curves are shown in Fig.~\ref{mag:sp6}(b).
The agreement is improved compared with case 1.
Therefore we will use the results of case 2 as our estimation of the model parameters for $\mathrm{Mo_{75}V_{20}}$ below.

We here comment on the magnetic impurities.
In the experiment of the spin ball $\mathrm{Mo_{72}V_{30}}$, the fresh sample gives almost the same susceptibility to the aged sample
and no Curie term due to magnetic impurities is observed.~\cite{Botor}
On the other hand, for the spin doughnut $\mathrm{Mo_{75}V_{20}}$ the difference of the aged and fresh samples is visible in the experiment.~\cite{Muller2}
As mentioned above, we have found that the difference between the two samples is resolved by subtracting the impurity contributions
and better agreement between the theory and experiment is obtained for the subtracted data than the raw data.
These facts may suggest that the main difference of the aged and fresh samples is in the impurity concentration, 
although the present authors have no idea about the mechanism of the aging process of $\mathrm{Mo_{75}V_{20}}$.

\begin{figure}[b]
\begin{center}
\includegraphics[width=.70\linewidth]{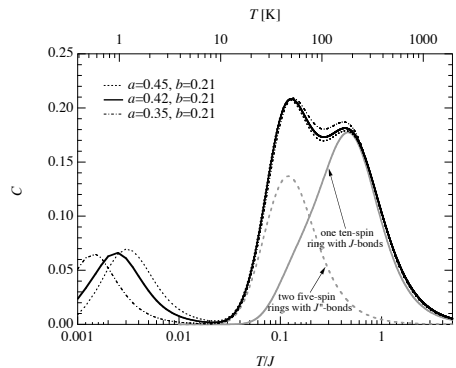}
\caption{Temperature dependence of $C(T)$ of the spin 
doughnut for $a=0.45$, 0.42 and 0.45, where $b=0.21$
is fixed. We used $J=388$ K to obtain the temperature in K.}
\label{Cv:sp}
\end{center}
\end{figure}

In order to get further insight into the magnetism of the spin doughnut, we study the temperature dependence of specific heat.
The calculated results are shown in Fig.~\ref{Cv:sp}, where we find a three-peak structure.
The specific heat for $T/J>0.03$ is approximately given by the sum of those of two five-spin rings and one ten-spin ring,
and thus the peak temperatures of the highest- and mid-temperature peaks scale $J$ and $J^{\prime\prime}$, respectively.
At low temperatures well below the mid-temperature peak position, two five-spin rings are expected to be in the doublet states, which is the origin of the low-temperature Curie law.
As seen in Fig.~\ref{Cv:sp}, the lowest-temperature peak position depends on the value of $a$, and it exists around $T=J^{\prime}/100$.
Below this temperature, the $J^{\prime}$-bonds get two five-spin rings and one ten-spin ring to be coupled each other to form a singlet ground state.

\begin{figure}[t]
\begin{center}
\includegraphics[width=.75\linewidth]{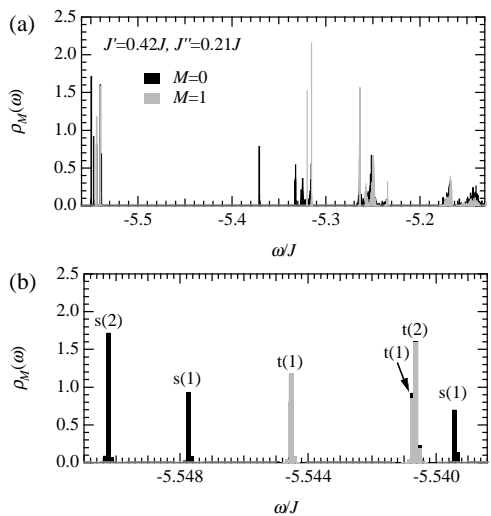}
\caption{Energy histograms $\rho_M(\omega)$ with 
total $S^z$ of $M=0$ and 1, where energy regions are 
chosen as (a) $\omega<\sim E_g+2 J^{\prime\prime}$ 
and (b) $\omega<\sim E_g+2 J^{\prime}/100$.
In the lower panel, s($l$) (t($l$)) denotes singlet (triplet) state with 
$l$-fold degeneracy.
}
\label{mag:sp7}
\end{center}
\end{figure}

It is instructive to look into the energy histograms $\rho_M(\omega)$ for the total $S^z$ of $M=0$ and 1.
For our model with $J^{\prime}=0.42J$ and $J^{\prime\prime}=0.21J$, the ground state energy is given by $E_g=-5.5503 J$.
Low energy parts of $\rho_M(\omega)$ are shown in Fig.~\ref{mag:sp7}(a) for $\omega-E_g< \sim 2 \times J^{\prime\prime}$
and in Fig.~\ref{mag:sp7}(b) for $\omega-E_g< \sim 2 \times J^{\prime}/100$, 
where the energy mesh, $\Delta\omega$, of the histogram is chosen as $\Delta\omega=10^{-5}J$.
We find in Fig.~\ref{mag:sp7}(a) that there is a group of states at $\omega\simeq E_g$.
Only the states in this group contribute to thermodynamics at low temperatures well below $J^{\prime\prime}$,
because energies of other excitations are higher than those by $0.17J\sim J^{\prime\prime}$.
As shown in Fig.~\ref{mag:sp7}(b), this group consists of four singlet and four triplet states.
Noting that the five-spin ring has two-fold degenerate doublet ground states and the ten-spin ring has a singlet ground state,
we interpret the four singlet and four triplet states as originating from direct products of ground states of the five- and ten-spin rings.

\begin{figure}[t]
\begin{center}
\includegraphics[width=.7\linewidth]{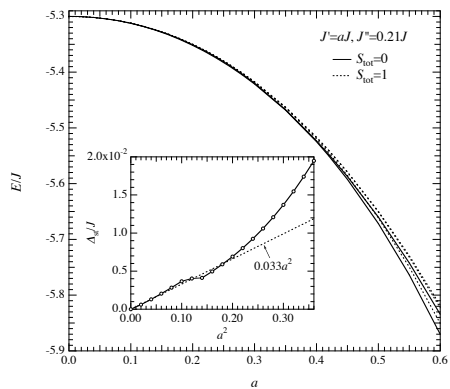}
\caption{Several low-energy eigenvalues for $S_{\rm{tot}}=0$ 
and 1, as a function of $a=J^{\prime}/J$.
The inset shows the singlet-triplet gap, 
$\Delta_{\rm{st}}$, versus $a^2$.
}\label{fig:beki}
\end{center}
\end{figure}

We here study the low energy singlet and triplet states as a function of $a=J^{\prime}/J$.
We calculate the lowest two eigenvalues in the singlet sector and the lowest three eigenvalues in the triplet sector
by the standard Lanczos method, and the results are shown in Fig.~\ref{fig:beki}.
As expected, we find that the energy eigenvalues reduced to the sum of ground state energies of five-spin and ten-spin rings as $a\rightarrow 0$,
and the energy splitting does not grow so much even for $a>b$.
It is natural to expect that this feature is originating from the geometrical structure of the $J^{\prime}$-bond, 
by which a spin on five-spin ring is connected to two spins on ten-spin ring.
In order to clarify this point, we plot the singlet-triplet gap $\Delta_{\rm{st}}$, 
which is defined by energy difference between the lowest eigenvalues in the singlet and triplet sectors, as a function of $a^2$ in the inset of Fig.~\ref{fig:beki}.
This plot gives the asymptotic form of $\Delta_{\rm{st}}/J\sim 0.033 a^2$.
Thus, the smallness of the energy splitting is coming from the fact of the lack of the first order term
and the smallness of perturbation coefficients for higher order terms.
We have to look into perturbation processes to understand the smallness of the perturbation coefficients.
However, such analysis remains as a future problem.


In summary, we have made a theoretical analysis of the susceptibility of the spin doughnut $\mathrm{Mo_{75}V_{20}}$,
where the impurity concentration has been estimated as 1.3\% (4.0\%) for the fresh (aged) sample
and the exchange parameters $J=388$ K, $J^{\prime}=163$ K and $J^{\prime\prime}=81$ K have been extracted from the experimental data.
On the basis of various calculations, we have discussed that the spin doughnut should be viewed as a weak coupling system of two five-spin rings and one ten-spin ring,
for which appearance of the low-temperature Curie law is quite natural because of the doublet states in the five-spin ring.
We have also predicted that the low-temperature Curie law ends around $T=1$ K, together with the lowest temperature peak of the specific heat.
It is desired to check those predictions experimentally, which judges the validity of our theory.
It is also interesting to observe the low-temperature specific heat under magnetic fields, 
because both the singlet and triplet states contribute the lowest temperature peak of the specific heat
and the energy differences are so small that modest values of the magnetic field can get the specific heat to change.

\acknowledgements
We have used a part of the code provided by H. Nishimori in TITPACK Ver.2. The authors are grateful to M. Nakane for useful discussions.

\end{document}